\documentclass{PoS}
\usepackage{subcaption}
\usepackage{amsmath}
\def\slashed#1{\kern+0.1em /\kern-0.65em #1}

\newcommand{\Imag}{\operatorname{Im}}

\newcommand{\tmop}[1]{\ensuremath{\operatorname{#1}}}

\title{Long distance part of $\epsilon_K$ from lattice QCD}

\ShortTitle{Long distance part of $\epsilon_K$ from lattice QCD}

\author{\speaker{Ziyuan Bai}%
       \thanks{A footnote may follow.}\\
      Columbia University in the city of New York\\
      E-mail: \email{zb2174@columbia.com}}

\abstract{We demonstrate the lattice QCD calculation of the long distance contribution
to $\epsilon_K$. Due to the singular, short-distance structure of
$\epsilon_K$, we must perform a short-distance subtraction and introduce a
corresponding low-energy constant determined from perturbation theory, which
we calculate at Next Leading Order (NLO).  We perform the calculation on a
$24^3 \times 64$ lattice with a pion mass of 329 MeV.  This work is a
complete calculation, which includes all connected and disconnected
diagrams.}

\FullConference{34th annual International Symposium on Lattice Field Theory\\
		 24-30 July 2016\\
		 University of Southampton, UK}

\begin{document}
\section{Introduction}

\begin{figure}[ht]
\centering
%\begin{minipage}{0.45\textwidth}
	\begin{tabular}{cc}
%\centering
		\includegraphics[width=0.45\textwidth]{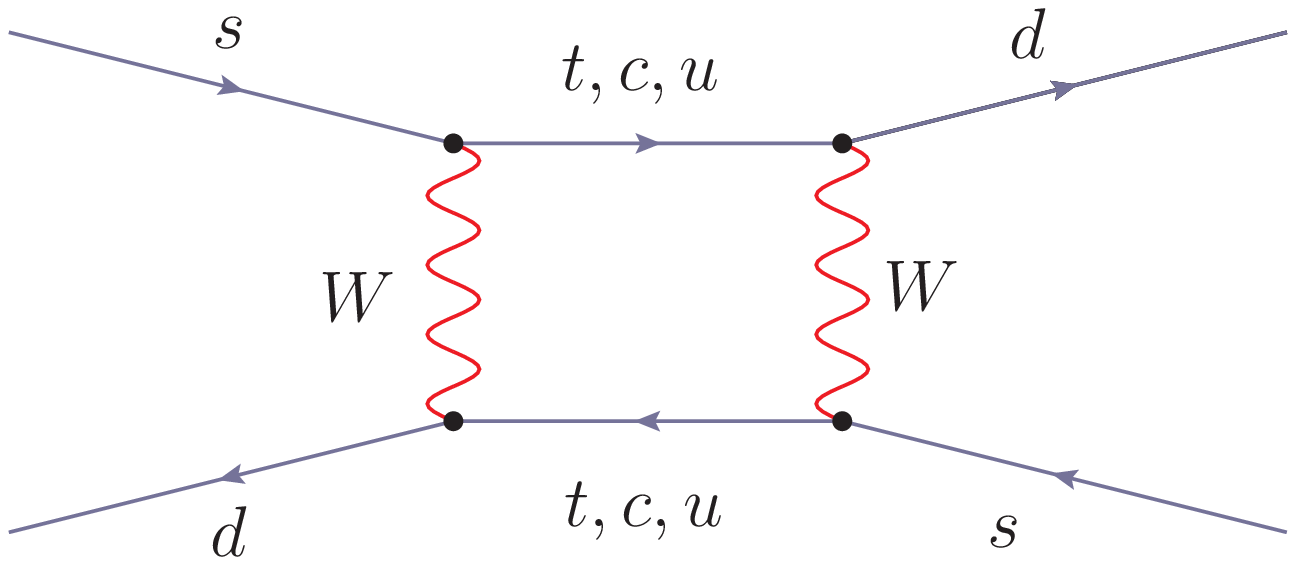} &
%\subcaption{Box} \label{fig:box}
%\end{minipage} \quad
%\begin{minipage}{0.45\textwidth}
%\centering
\includegraphics[width=0.45\textwidth]{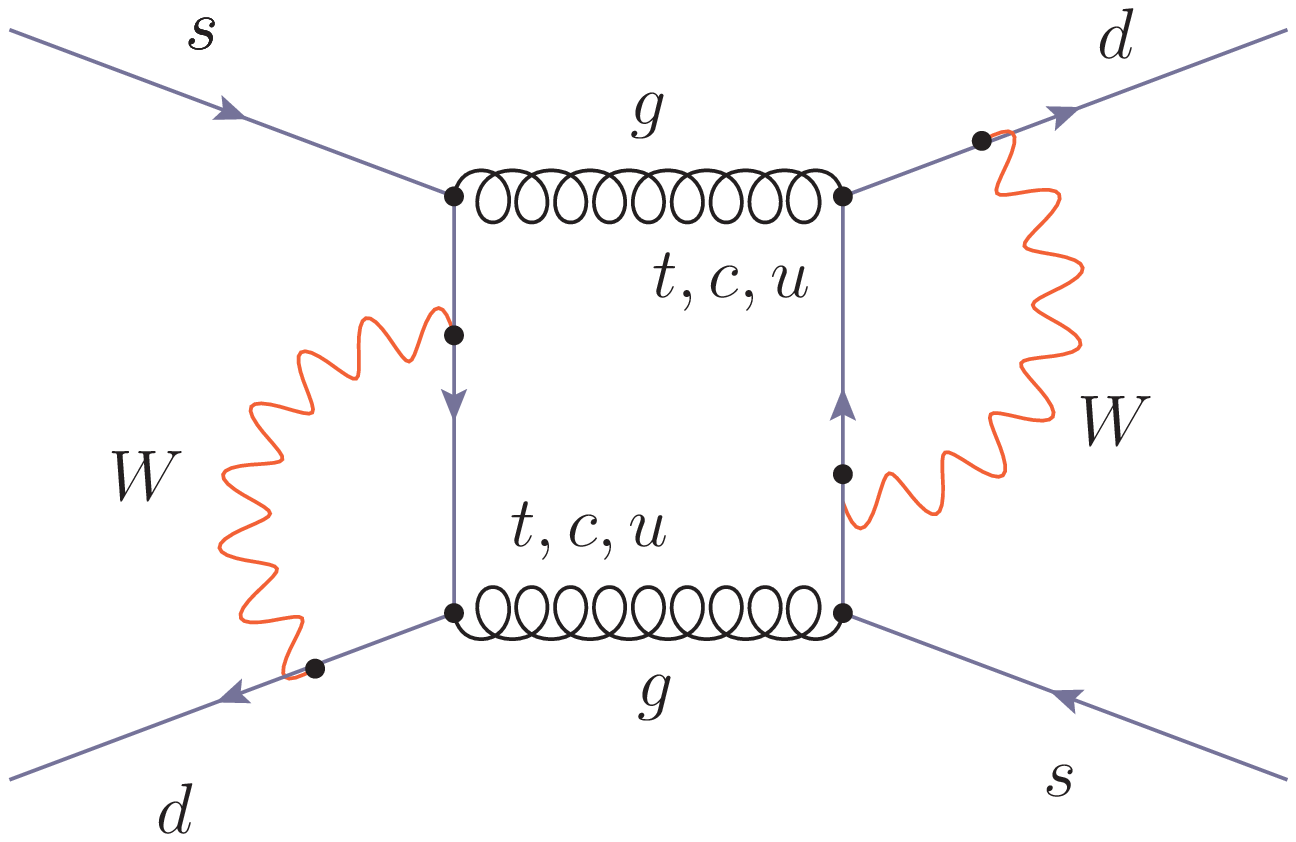} \\
box & disconnected\\
\end{tabular}
%\subcaption{Disconnected}\label{fig:disconnected}
%\end{minipage}
\caption{Two type of diagrams contribute to the $\Delta S = 2$ process.}
\label{WW_diagram}
\end{figure}

The $K^0 - \overline{K^0}$ mixing parameter $\epsilon_K$, with an 
experimental value of $2.228(11)\times10^{-3}$, provides a measure of 
indirect CP violation in $K^0 - \overline{K^0}$ mixing. This amplitude is caused by $\Delta S = 2$
weak interaction and is second order in the Fermi constant $G_F$. Two types of diagrams which contribute to
this process are shown in Fig.~\ref{WW_diagram}.
The usual estimation from the standard
Model of $\epsilon_K$ involves only the short distance contribution, calculated
by first integrating out the W boson and top quark, resulting in a bi-local effective weak Hamiltonian. One
then integrates out the charm quark 
and treats the bi-local weak Hamiltonian as a local operator multiplied by a Wilson 
coefficient determined
in perturbation theory. We have seen a 3.6(2) $\sigma$ tension between the experimental value if we 
use the exclusive $V_{cb}$ while the tension goes away if use choose to use 
inclusive $V_{cb}$ \cite{Bailey:2015wta}. 

To understand the Standard Model contribution to $\epsilon_K$, a calculation with the long distance
part correctly controlled is needed. The previous estimation from 
Chiral Perturbation Theory is a few percent ~\cite{Buras:2010pza}. 
The calculation including the long distance part can be done using lattice QCD to calculate
the bi-local part of the $K^0 - \overline{K^0}$ mixing matrix element
and correcting the short distance divergence 
of the lattice calculation by performing a perturbative matching. In our first attempt \cite{NHC:2015PoS},
we have included the type 1\&2 diagrams in our analysis and 
performed the perturbative matching to LO, which gave us a quite large systematic error because
the NLO correction can be 50\%. In this calculation, we have included all the diagrams and use
a NLO perturbative matching with the effective Hamiltonian instead of matching to the box
diagram in full theory.

\section{Effective $\Delta S = 2$ Hamiltonian}

The parameter $\epsilon_K$ can be calculated using Eq.~\ref{epsK}, where $\phi_{\epsilon} = 43.52 \pm 0.005^\circ$,
and $\xi$ is the ratio of imaginary part of $K^0 \rightarrow \pi\pi$ 
matrix element $A_0$ to its real part. $\Delta M_K$
is the $K_L - K_S$ mass difference which has the experimental value $3.483(6)\times 10^{-12}$ MeV.
We will have to calculate the imaginary part of the kaon mixing matrix element 
$M_{\bar{0}0} = \frac{\langle \overline{K^0} | H_W^{\Delta S = 2} | K^0 \rangle}{2 m_K} $ from the Standard Model.
The conventional method to calculate $M_{\bar{0}0}$ involves first integrating out W boson and top quark, 
and then integrating out charm to get the weak Hamiltonian in Eq.~\ref{Heff}. 
In writing out the effective Hamiltonian, we use a slightly different
approach: we use the unitary condition of CKM matrix to convert the sum over three up-type flavors 
in the internal quark lines into 
three pieces: $(u-c)\times (u-c)$, $(t-c) \times (t-c)$, and $(t-c) \times (u-c)$. This is different from subtracting 
the up quark in each piece which is commonly done. With our choice, we only have to consider the term
that has $\lambda_u \lambda_t$ and involves the combination
$(t-c)\times(u-c)$, because the other two terms are either purely a short-distance contribution,
or do not receive an imaginary part from the CKM matrix.

\begin{eqnarray}
\epsilon_K = \exp{i\phi_{\epsilon}} \sin{\phi_{\epsilon}} \left( 
\frac{-\Imag{M_{\bar{0}0}}}{\Delta M_K} + \xi \right) 
\label{epsK}
\end{eqnarray}

\begin{eqnarray}
	H_{\tmop{eff}}^{\Delta S=2} & = & \frac{G_{F}^{2}}{16 \pi^{2}} M_{W}^{2} [
  \lambda_{u}^{2} \eta'_{1} S'_{0} ( x_{c} ) + \lambda_{t}^{2} \eta'_{2}
  S'_{0} ( x_{t} ) +2 \lambda_{u} \lambda_{t} \eta'_{3} S'_{0} ( x_{c} ,x_{t}
) ]  Z(\mu) Q_{\tmop{LL}}
	\label{Heff}
\end{eqnarray}

To perform a long distance calculation, we do not integrate out the charm quark in $H_{eff,ut}^{\Delta S = 2}$,
where to $ut$ means the term contains a $\lambda_u \lambda_t$ factor. 
Instead, we calculate in the four flavor theory
where we can expand the weak Hamiltonian as a sum over products of pairs of 
$\Delta S = 1$ operators, as shown in Eq.~\ref{Heff_ut}.

\begin{eqnarray}
	H_{eff,ut}^{\Delta S = 2} &=& \frac{G_F^2}{2} \lambda_u \lambda_t \left[\sum_{i=1}^{2} 
	\sum_{j=1}^{6} C_i(\mu) 
	C_j(\mu) [Q_i Q_j ] \label{Heff_ut}
+ C_7(\mu) O_{LL} \right],
\end{eqnarray}

In Eq. ~\ref{Heff_ut}, $Q_{1,2}$ are the 
current-current operators and $Q_{3,4,5,6}$ are the QCD penguin operators, whose definitions
can be found in \cite{Buras}. The 
structure of the product of operators $[Q_i Q_j]$ 
can be found in Eq.12.40 - 12.41 of \cite{Buras}, where the
only difference is that we have a factor of $c-u$ propagator difference times a $c$ propagator,
instead of a $u-c$ times a $u$ propagator. 
The operator $O_{LL} = (\bar{s} d)_{V-A} (\bar{s} d)_{V-A}$ is the local 
operator which changes the strangeness by 2. The product 
$[Q_i Q_j]$ will be logarithmically divergent when these two
operators become close to each other because we do not
have a GIM cancellation in both of the internal quark
lines. This divergence will be absorbed by the Wilson coefficient of the operator $Q_{LL}$.
In the lattice calculation, this quantity will also be ultra-violet divergent with the high energy cutoff
determined by the inverse lattice spacing $1/a$. This divergence is unphysical and we have to remove
it by subtracting the $O_{LL}$ operator multiplied by a coefficient matched to the continuum. 
This matching will require a non-perturbative calculation 
on the lattice and a corresponding perturbative calculation in the continuum. This is done in an 
Regularization Independent (RI) intermediate scheme. To define our RI scheme, we write our RI operator
in terms of both the $\overline{MS}$ operator (Eq.~\ref{RI_op1}) and the lattice operator (Eq.~\ref{RI_op2}),
we then impose the RI condition that these operators vanish when inserted in a Landau gauge-fixed 
Green's function evaluated at 
off-shell momenta with 
a scale $\mu_{RI} \gg \Lambda_{QCD}$ . We can find 
both the coefficients $X^{i,j}$ and $Y^{i,j}$ in Eq.~\ref{RI_op1} and Eq.~\ref{RI_op2}. 
This procedure is
similar to what we have done in the $K \rightarrow \pi \nu \bar{\nu}$ calculation \cite{Christ:2016eae},
but here we write our formula in a fashion in which we absorb some Wilson coefficients into $X$ and $Y$.

%\small{
%\begin{eqnarray}
%	\left[Q_i Q_j\right]^{RI}(\mu_{RI}) &=&Z_i^{lat \rightarrow RI}(\mu_{RI},a)
%	Z_j^{lat \rightarrow RI}(\mu_{RI},a) 
%	[Q_iQ_j]^{lat} - X^{i,j}(\mu_{RI},a) 
%	Z^{lat \rightarrow RI}
%	(\mu_{RI},a) O^{lat}_{LL}  \label{RI_op1} \\
%	%[Q_i Q_j]^{RI}(\mu_{RI}) = Z_i^{\overline{MS} \rightarrow RI}(\mu,\mu_{RI})  \label{RI_op2}
%	\left[Q_i Q_j\right]^{RI}(\mu_{RI}) &=& Z_i^{\overline{MS} \rightarrow RI}(\mu,\mu_{RI})  
%	Z_j^{\overline{MS} \rightarrow RI} (\mu_{RI})  
%	[Q_i Q_j]^{\overline{MS}}  - Y^{i,j}(\mu, \mu_{RI})Z^{\overline{MS}
%\rightarrow RI}(\mu, \mu_{RI})
%	O_{LL}^{\overline{MS}}  \label{RI_op2}
%\end{eqnarray}
%}

{\small
\begin{eqnarray}
	\left[Q_i Q_j\right]^{RI}(\mu_{RI}) &=&Z_i^{lat \rightarrow RI}(\mu_{RI},a)
	Z_j^{lat \rightarrow RI}(\mu_{RI},a)  \left\{
	[Q_iQ_j]^{lat} - X^{i,j}(\mu_{RI},a) 
 O^{lat}_{LL} \right\} \label{RI_op1}, \\
	%[Q_i Q_j]^{RI}(\mu_{RI}) = Z_i^{\overline{MS} \rightarrow RI}(\mu,\mu_{RI})  \label{RI_op2}
	\left[Q_i Q_j\right]^{RI}(\mu_{RI}) &=& Z_i^{\overline{MS} \rightarrow RI}(\mu,\mu_{RI})  
Z_j^{\overline{MS} \rightarrow RI} (\mu,\mu_{RI})  \left\{
	[Q_i Q_j]^{\overline{MS}}  - Y^{i,j}(\mu, \mu_{RI})
O_{LL}^{\overline{MS}}  \right\}. \label{RI_op2} 
\end{eqnarray}
}

In Eq.~\ref{RI_op2}, we will insert external momentum at scale $\mu_{RI}$ to 
evaluate the quantity $Y^{i,j}$. The conventional SM calculation of $\epsilon_K$ 
includes integrating out the charm
quark, which is a similar calculation and is usually done at zero external momentum.
Therefore, we define $\Delta Y^{i,j}(\mu, \mu_{RI})$ as the difference between
$Y^{i,j}(\mu, \mu_{RI})$ evaluated at our $\mu_{RI}$ external momentum and the same quantity evaluated 
at zero external momentum.
We can then write the total $\Delta S = 2$ weak Hamiltonian in a very clean way:

\begin{eqnarray}
\nonumber
H_{eff,ut}^{\Delta S = 2} &=& \sum_{i=1}^{2} \sum_{j=1}^{6}    \left\{
%	C_i^{\overline{MS}}(\mu)C_j^{\overline{MS}}(\mu)
%	Z_i^{lat \rightarrow \overline{MS}} Z_j^{lat \rightarrow \overline{MS}}
C_i^{lat}(\mu)C_j^{lat}(\mu)
	\left([Q_i Q_j]^{lat}
	- X^{i,j}(\mu_{RI}) O^{lat}_{LL}\right) \right. \\\nonumber
	&& + C_i^{\overline{MS}}(\mu)C_j^{\overline{MS}}(\mu)
	\left[Y_{\overline{MS}}^{i,j}(\mu, \mu_{RI}) - Y_{\overline{MS}}^{i,j}(\mu, 0)\right]
	Z^{lat\rightarrow \overline{MS}} O^{lat}_{LL} \\
	&&\left. + \left[ C_i^{\overline{MS}}(\mu)C_j^{\overline{MS}}(\mu) 
Y_{\overline{MS}}^{i,j}(\mu, 0) + C_7^{\overline{MS}}(\mu) \right] 
Z^{lat\rightarrow \overline{MS}} O^{lat}_{LL}   \right\} .
 \label{masterEq}
\end{eqnarray}

\begin{figure}[ht]
	\centering
	\includegraphics[width=0.8\textwidth]{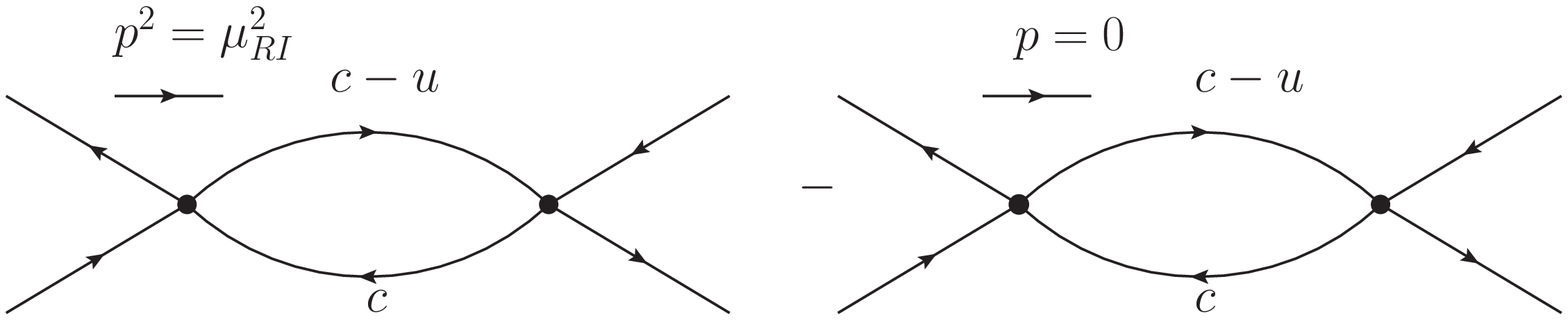}
	\caption{Illustration of the perturbative calculation of $\Delta Y_{\overline{MS}}^{i,j}( \mu_{RI})$}
	\label{delta_Y}
\end{figure}

The first line of Eq.~\ref{masterEq} involves the lattice operators and the coefficients $X^{i,j}$
determined from non-perturbative renormalization (NPR). We call this term the ``contribution
below $\mu_{RI}$'', which includes the long distance part,
and use $\Imag{M_{\bar{0}0}^{ut,RI}(\mu_{RI})}$ to denote its contribution to the kaon mixing matrix element.
The second line involves coefficient 
$	\Delta Y_{\overline{MS}}^{i,j}(\mu_{RI}) = Y_{\overline{MS}}^{i,j}(\mu, \mu_{RI}) -
Y_{\overline{MS}}^{i,j}(\mu, 0)$
calculated from perturbation theory, and we call this term
``perturbative RI to $\overline{MS}$ correction'', $\Imag{M_{\bar{0}0}^{ut,RI\rightarrow \overline{MS}}(\mu_{RI})}$.
The last term is the result of conventional standard model calculation.
We will label it the ``conventional short distance result''.
The combination of the first two terms is our ``long distance correction'' to the SM calculation of 
$\epsilon_K$. It should be independent of the RI scale $\mu_{RI}$ we introduced. The 
perturbative calculation 
of $\Delta Y_{\overline{MS}}^{i,j}(\mu_{RI})$ is illustrated in Fig \ref{delta_Y}. We note that
this calculation is both infra-red and ultra-violet convergent at NLO, because of the non-zero charm
quark mass 
and the subtraction between two logarithmically UV divergent diagrams. The NPR calculation of $X^{i,j}$ is the same
as what we have done in \cite{NHC:2015PoS}.

\section{Lattice calculation and results}

\begin{figure}[ht]
	\centering
	\begin{tabular}{ccc}
		\includegraphics[width=0.35\textwidth]{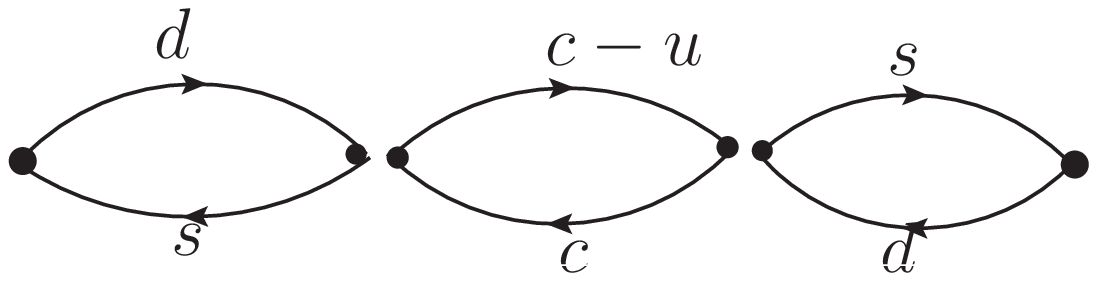} &
		\includegraphics[width=0.3\textwidth]{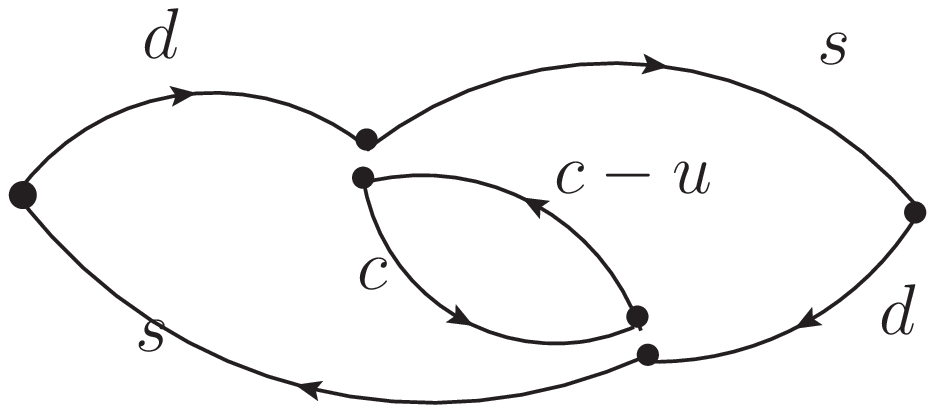} &
		\includegraphics[width=0.3\textwidth]{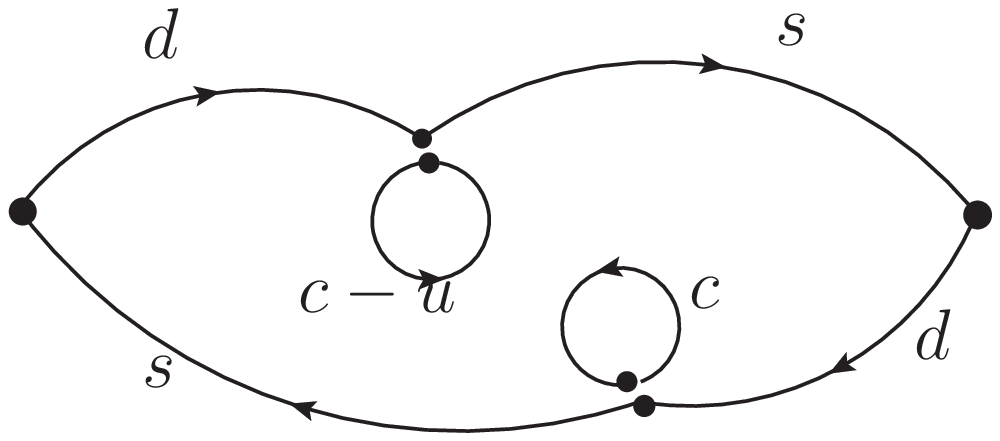} \\
	\end{tabular}
	\begin{tabular}{cc}
		\includegraphics[width=0.4\textwidth]{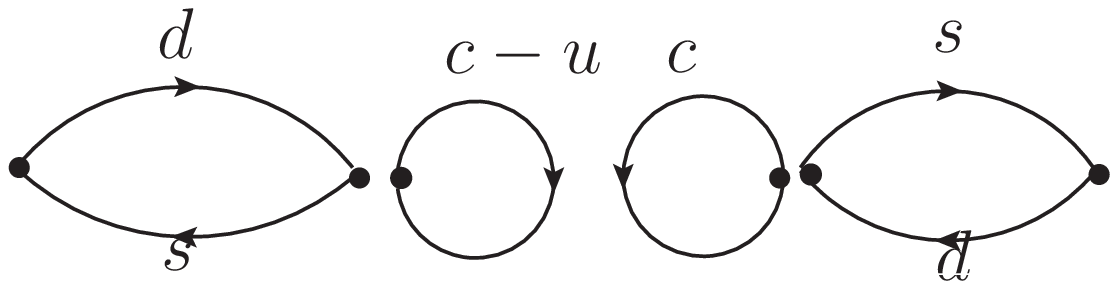} &
		\includegraphics[width=0.4\textwidth]{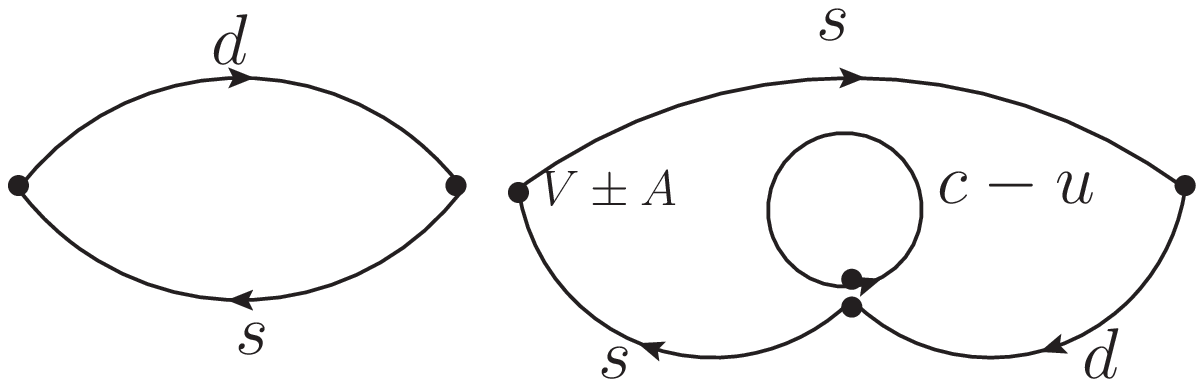} \\
	\end{tabular}
	\caption{Five types of contractions appearing in this calculation. On the top: type 1,2,3 diagrams and on the bottom,
	type 4,5 diagrams, listed from left to right.}
	\label{5diag}
\end{figure}

We have carried out this calculation on our $24^3\times 64$ Iwasaki gauge ensemble with an inverse lattice spacing 1.78 GeV and
a pion mass 338 MeV. The kaon mass is 591 MeV which is below the two pion energy. The
charm quark is unphysical with a mass of 968 MeV. We use the method we 
introduced in the unphysucal $\Delta M_K$ calculation \cite{Yu:klks16}, integrating over the position of the two
weak Hamiltonians, and subtracting the contribution of all intermediate 
states lighter than the kaon, which in this case are the
vacuum and the single pion state. We have contractions between two current-current operators or 
between one current-current operator
and one QCD penguin operator. We have 5 types of four point diagrams to compute on the lattice, where the type 5
diagram will only appear when we have a penguin operator. All the different
contractions are shown in Fig~\ref{5diag}.
The type 1,2,3,4 diagrams are shown with two current-current operators and have $c\times(c-u)$ 
propagator combinations in the internal loop. For the diagram with one current-current
operator and one penguin operator, we will have a $cc-uu $ difference for the 
propagators in the internal loop and $(c-u) \times (u+d+s+c)$ in the
self loops of type 3\&4 diagrams. The type 5 diagram shown has a vertex of $\bar{s}d \bar{s} s$ on the left, and
we also have a similar diagram with a $\bar{s}d \bar{d} d$ vertex. 

We have used Coulomb gauge fixed wall source for the two kaons, and we solve for a source at each time slice so
we can perform a time translation for all the contractions we calculate.
In the type 1\&2 diagrams, we have used a point source at each time slice for one of the two weak vertices, while
the other weak vertex is summed over the spacial volume. 
For the type 3\&4 diagrams, the self-loop is computed with an all-to-all propagator whose 
high-mode part is obtained from a sum of 60 random, space-time volume sources, 
while the low mode part is constructed from 
450 eigenvectors obtained from the Lanczos algorithm.
In the type 5 diagrams, we use the same point source propagator as in the type 1\&2 diagrams
for the self-loop arising from the current-current operators, and the vertex of the penguin operators is summed over
the spacial volume. In our first measurement, we had
problems with our random number generator so we measured the type 3\&4 diagrams again. In this analysis we used 
75 configurations with a correct random number generator for type 3\&4 diagrams and 150 configurations
for type 1\&2\&5 diagrams which do not require random numbers.

Our lattice calculation of $\Imag{M_{\bar{0}0}^{ut,RI}}$, which corresponds to the contribution from those parts of 
$H_W$ in the first line of Eq.~\ref{masterEq}, is shown in Table \ref{sep_result}. We have listed the result
from the different operator combinations and the Wilson coefficients have been included and the divergence
has been removed by including the $X^{i,j}(\mu_{RI})$ terms. We can see that the current-current operators
give the largest contributions because of their larger Wilson coefficients. We have shown the lattice Wilson
coefficient for each operator in Table~\ref{WilsonCoef}. 
%We note that the Eq.~\ref{masterEq} is misleading
%in the sense that the $Z^{lat->\overline{MS}}$ is mixing matrix that involves non-zeros off diagonal element,
%and the lattice Wilson coefficient should be calculated using
%$C_i^{lat} = C_j^{\overline{MS}} Z_{ji}^{lat->\overline{MS}}$.

\begin{table}[ht]
	\centering
	\begin{tabular}{c|c|c|c|c|c}\hline
		$Q_1 Q_1$ & $Q_1 Q_2$ & $Q_1 Q_3$ & $Q_1 Q_4$ & $Q_1 Q_5$ & $Q_1 Q_6$ \\\hline
		& $Q_2 Q_2$ & $Q_2 Q_3$ & $Q_2 Q_4$ & $Q_2 Q_5$ & $Q_2 Q_6$ \\\hline
   0.4436(321)  & -0.2540(1364)  & -0.0696(112) &   0.0004(206)  &  0.0228(173)  & -0.1701(1479)\\\hline
        0(  0)  &  1.6991(2232)  & -0.0284(201) &  -0.1405(419)  &  0.0310(448)  & -0.2220(3850)\\\hline
	\end{tabular}
	\caption{Contribution to $\Imag{M_{\bar{0}0}^{ld}}$ by different 
	operator combination, all 5 types of diagrams have been included. We have choosed $\mu_{RI} = 1.92$ GeV.
From correlated fit with fitting range 8:16.}
	\label{sep_result}
\end{table}

\begin{table}[ht]
	\centering
	\begin{tabular}{c|c|c|c|c|c}\hline
	$C_1$ & $C_2$ & $C_3$ & $C_4$ & $C_5$ & $C_6$ \\\hline
		0.2373(1) & 0.6885(1) & 0.0113(6)&  0.0213(7) & 0.0085(6) & 0.0256(6)\\\hline
	\end{tabular}
	\caption{Lattice Wilson coefficient, calculated using $(\gamma_\mu, \gamma_\mu)$ intermediate scheme 
	in the NPR step, and the $\overline{MS}$ Wilson coefficient are obtained with $\mu = 2.15$ GeV.}
	\label{WilsonCoef}
\end{table}

The sum over all terms in Table ~\ref{sep_result} is $\Imag{M_{\bar{0}0}}^{ut,RI} = -1.49(69)$, and its
total contribution to $\epsilon_K$ is $|\epsilon_K^{ut,RI}| = 3.0(14)\times10^{-4}$. To get our final correction
to $\epsilon_K$, we must include the second line of Eq.~\ref{masterEq}, which we call
$\Imag{M_{\bar{0}0}}^{ut,RI\rightarrow \overline{MS}}$. This result is summarized in 
Table~\ref{final}, and we have five different intermediate values for $\mu_{RI}$ to test the consistency of
our result. In calculating $\Imag{M_{\bar{0}0}}^{ut,RI\rightarrow \overline{MS}}$, we have to specify the
charm quark mass to be used in the perturbative calculation. Because this is a NLO calculation and our
answer is accurate to order $\mathcal{O}(\alpha_s \ln{\mu/M_W)}$. The charm quark mass has a dependence
on the scale $\mu_{\overline{MS}}$ or $\mu_{RI}$, which is of order  $\alpha_s$, allowing 
us to ignore the scale dependence of
charm quark mass and use a mass of 968 MeV, which is same as our lattice input converted to $\overline{MS}$
at 2 GeV (We have used $Z_m^{lat\rightarrow \overline{MS}} (2 \mbox{ GeV}) = 1.498$ from \cite{Aoki:2010dy}).
The $\Imag{M_{\bar{0}0}}^{ut,RI\rightarrow \overline{MS}}$ contribution also involves the calculation of the kaon bag
parameter $B_K$, which also has a scale dependence on $\mu_{\overline{MS}}$, and we have used the $B_K$ 
evaluated at 2.15 GeV. We can see some dependence on $\mu_{RI}$ in Table \ref{final}.
We found that the matching from the RI scheme to $\overline{MS}$ is successful that 
the results in the last column, which is our long distance correction to $\epsilon_K$, depend
very little on the intermediate scale $\mu_{RI}$ (when the $\mu_{RI}$ is larger than 2 GeV).
If we see a discrepancy between different $\mu_{RI}$, then it might indicate we'll need a NNLO 
$\Imag{M_{\bar{0}0}}^{ut,RI\rightarrow \overline{MS}}$
to get a more consistent result. In such a NNLO matching,
we must take the scale dependence of $m_c$ and $B_K$ into consideration.

\begin{table}[ht]
	\centering
	\begin{tabular}{c|c|c|c|c}\hline
	$\mu_{RI}$ & $\Imag{M_{\bar{0}0}}^{ut,RI}$ & $\Imag{M_{\bar{0}0}}^{ut,RI \rightarrow \overline{MS}}$ &
		$\Imag{M_{\bar{0}0}}^{ut,ld\, corr}$ & contribution to $\epsilon_K$ \\\hline
		1.54 & -1.12(52) & 0.352  & -0.77(52) & $0.151(102)\times10^{-3}$ \\\hline
		1.92 & -1.31(51) & 0.476 & -0.83(51)  & $0.164(100)\times10^{-3}$  \\\hline
		2.11 & -1.40(52) & 0.537  & -0.86(52) & $0.170(102)\times10^{-3}$\\\hline
		2.31 & -1.47(51) &  0.599 & -0.87(51) & $0.171(100)\times10^{-3}$\\\hline
		2.56 & -1.55(51) &  0.674 & -0.87(51) & $0.172(100)\times10^{-3}$\\\hline
	\end{tabular}
	\caption{Long distance correction to $\epsilon_K$ as we vary the intermediate scale $\mu_{RI}$.
The fourth column is the sum of the results in the previous two column and the last column is the corresponding
	contribution to $\epsilon_K$. }
	\label{final}
\end{table}

\section{Conclusion}

In this exploratory calculation with unphysical charm and light quark mass,
we find a long distance correction to $\epsilon_K$ of $0.170(100)\times10^{-3}$(using the
result of $\mu_{RI} = 2.11$ GeV), which is about 
8\% of the total $\epsilon_K$ experimental value. This is the correction
that should be added to the Standard Model result of $\epsilon_K$ that is usually presented.
In our method, instead of integrating the charm quark at zero external momentum 
and treating the bi-local operator as a local operator,  we use a momentum $\mu_{RI}$ that can be 
chosen far above the charm quark mass (of course, this scale is limited by the inverse lattice spacing $1/a$)
in this perturbative calculation.
Therefore we have much better control of the systematic error in perturbation theory 
, and are able to evaluate the low-energy, non-perturbative part in our lattice calculation. Our 
matching to perturbation theory calculation for $\Delta Y_{\overline{MS}}(\mu_{RI})$ 
is done at NLO. To do this in NNLO,
we must perform a two loop calculation which is significantly more difficult than our current one-loop
perturbative calculation. The correction from NLO to NNLO to the
Standard Model estimation to $\epsilon_K$ is only a few percent, but it may 
change our conclusion because the size of our long distance correction to $\epsilon_K$ is also at
the $10\%$ level.
%but the dependence on $\mu_{RI}$
%in Table \ref{final} may suggest the size of a NNLO contribution.

We must note that this is not a physical calculation, in that we have a heavier than physical light quark mass 
which corresponds to a pion mass of 329 MeV. 
We also have an unphysical charm quark mass of 968 MeV, because we cannot go to a larger charm 
quark mass in our Iwasaki lattice with $1/a = 1.78$ GeV and a domain wall fermion action. 
A physical calculation with physical quark masses
on a finer lattice will provide more realistic information than the current one.

\bibliographystyle{plain}
\bibliography{reference}
%\begin{thebibliography}{99}
% \bibitem{reference.bib} 
%\end{thebibliography}

\end{document}